\documentclass[12pt]{iopart}
\usepackage{graphicx}

\begin{document}
\title{Two-bit Deutsch-Jozsa algorithm using an atomic ensemble}

\author{Shubhrangshu Dasgupta$^1$}
\address{$^1$Department of Chemistry, University of Southern California, Los Angeles, CA 90089, USA}
\author{G. S. Agarwal$^2$\footnote{On leave of absence from Physical Research Laboratory, Navrangpura, Ahmedabad - 380 009, India}}
\address{$^2$Department of Physics, Oklahoma State University, Stillwater, OK - 74078, USA}
\eads{\mailto{shubhrad@usc.edu},\mailto{agirish@okstate.edu}}
\date{\today}
\begin{abstract}
The first optical proposal for the realization of the two-bit version of the Deutsch-Jozsa
algorithm [D. Deutsch and R. Jozsa, Proc. R. Soc. London A {\bf 493}, 553 (1992)] is presented. The
proposal uses Stark shifts in an ensemble of atoms and degenerate sources of photons. The photons
interact dispersively with an atomic ensemble, leading to an effective Hamiltonian in atom-field
basis, which is useful for performing the required two-qubit operations. Combining these with a set
of one-qubit operations, the algorithm can be implemented. A discussion of the experimental
feasibility of the proposal is given.
\end{abstract}
\maketitle

It is known that the quantum computers are able to perform certain
specific jobs like factorization, searching a database etc. much
faster than its classical counterparts. Several quantum algorithms
have been developed to demonstrate the power of quantum computers.
For example, Grover has proposed an algorithm to search a quantum
state from an unsorted database of $2^n$ states \cite{grover}.
This requires $O(2^{n/2})$ repetitions of certain unitary
operations, demonstrating quadratic speed-up than the classical
computers. Deutsch-Jozsa algorithm \cite{dj} can be used to
identify a certain function of $n$ binary variables as constant or
balanced by a single enquiry, whereas classically it would require
up to $2^{n-1}+1$ enquiries. The key of this enormous speed of
quantum computers lies in the so-called quantum parallelism, which
enables one to apply the same unitary operation simultaneously on
a number of basis states.

There have been several proposals and experiments on implementation of these algorithms. For
example, Grover's algorithm has been implemented using nuclear magnetic resonance (NMR) in bulk
systems \cite{groverNMR}. NMR \cite{djNMR1,djNMR2}, ion trap \cite{djION}, and linear optical
\cite{mohseni} techniques have been employed to implement the Deutsch-Jozsa algorithm. Recently, an
optical scheme to realize the one-bit version of the Deutsch-Jozsa algorithm was proposed
\cite{chandan}. To date, most of the works on Deutsch-Jozsa algorithm is for the case of a single
qubit and we need to realize this algorithm for larger number of qubits, so that the full power of
this algorithm can be realized. In this paper, we show how this can be achieved using Stark shifts
in an ensemble of atoms. In this model, two freely propagating photonic qubits encoded in their
polarization modes interact dispersively with an atomic ensemble. This leads to an effective
Hamiltonian which along with several one-qubit operations enables us to implement the Deutsch-Jozsa
algorithm for two qubits. We further discuss the experimental feasibility of our scheme using
current technology.

We start with the main features of the algorithm. This algorithm helps to identify whether a
bivalued function $f_i(x)$ (which can take only the values 0 or 1) of a variable $x$ is constant or
balanced. Here $x$ is the decimal equivalent of the bit string $(x_1,x_2,...,x_n)$ for $n$ bits,
i.e., $x=\sum_{i=1}^nx_i2^{n-i}$. Thus, $x$ can take any of the $2^n$ values between 0 and $2^n-1$,
for different binary combinations. If for half the values of $x$, the function takes the value $0$
and for the other half, it takes the value 1, then the function is called balanced. The function
will be called constant if it assumes either the value 0 or 1 for all values of $x$. In classical
sense, it is easy to verify that at most $2^{n-1}+1$ evaluations of the function $f_i(x)$ for
different values of $x$ are necessary to determine whether the function is constant or balanced
[see Table \ref{table1}]. The Deutsch-Jozsa algorithm, on the contrary, requires only a single
``evaluation" of the function for the same. To demonstrate such an immense power of this algorithm,
we show in Fig. 1 the basic circuit \cite{chuang} to perform the $n$-bit version of the
Deutsch-Jozsa algorithm. It can be shown that if all the qubits of the system A are in $|0\rangle$
state, then the function $f_i(x)$ is a constant function. But if at least one of these $n$ qubits
is in the state $|1\rangle$, then the function can be identified as a balanced function. In what
follows, we will demonstrate the two-bit version of the Deutsch-Jozsa algorithm in an ensemble of
atoms interacting with a quantized field.

\begin{table}
\caption{\label{table1}Different possible functions showing their characterizations for $n=2$. Here
$x$ is the decimal equivalent of the bit string $(x_1,x_2)$, e.g., $x=2$ for two-bit input (1,0).}
\begin{center}
\begin{tabular}{c|cc|cccccc}
\hline \hline
Input&\multicolumn{2}{c|}{Constant}&\multicolumn{6}{c}{Balanced}\\
\hline
$x$&$f_1(x)$ & $f_2(x)$ & $f_3(x)$ & $f_4(x)$ & $f_5(x)$ & $f_6(x)$ & $f_7(x)$ & $f_8(x)$ \\
\hline
0 & 0 & 1 & 0 & 1 & 0 & 1 & 0 & 1\\
1 & 0 & 1 & 0 & 1 & 1 & 0 & 1 & 0\\
2 & 0 & 1 & 1 & 0 & 0 & 1 & 1 & 0\\
3 & 0 & 1 & 1 & 0 & 1 & 0 & 0 & 1\\
\hline \hline
\end{tabular}
\end{center}
\end{table}

\begin{figure}
\centerline{\scalebox{0.6}{\includegraphics{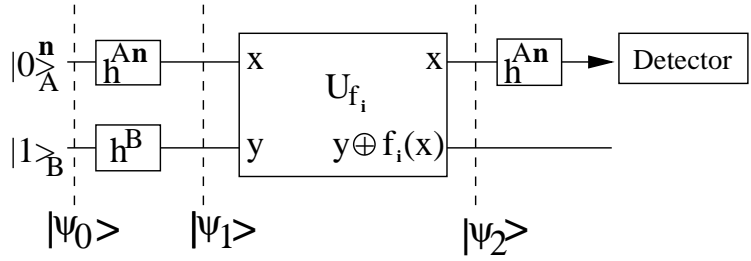}}} \caption{\label{fig1}Basic circuit to
perform the Deutsch-Jozsa algorithm for $n$ qubits. All the $n$ qubits in the system A are
initialized to the state $|0\rangle$, while the qubit in the ancillary system B is initialized to
the state $|1\rangle$. $h^A$, $h^B$ - Hadamard transformations of the qubits of the systems A and
B, $U_{f_i}$ - unitary operation operating on the system A+B, $|\psi_0\rangle$ etc. - states of the
system A+B, $x\in \{0, \ldots , 2^n-1\}$, $y\in \{0,1\}$. The detector measures states of each
qubit of the system A in $(|0\rangle,|1\rangle)$ basis.}
\end{figure}

\begin{figure}
\centerline{\scalebox{0.9}{\includegraphics{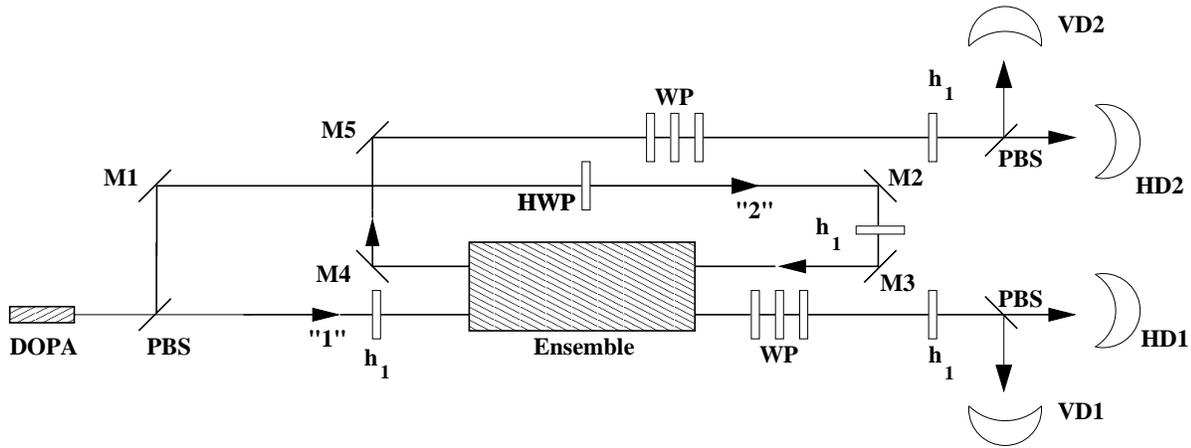}}} \caption{\label{fig3} A possible
experimental set-up for the two-bit Deutsch-Jozsa algorithm. DOPA - degenerate optical parametric
amplifier, PBS - polarization beam splitter, M1 etc. - lossless mirrors, HD1, HD2, VD1, VD2 -
single-photon detectors for horizontal (H) and vertical (V) polarizations, HWP - half-wave plate,
h$_1$ - sequence of quarter-wave and half-wave plates to perform the operation $h_1$ for photons,
WP - sequence of quarter-wave and half-wave plates to perform the operations $h'$ and $h''$ for
photons. The Hadamard rotations for the atomic ensemble are not shown.}
\end{figure}

In order to implement the Deutsch-Jozsa algorithm for two qubits, we need two distinguishable
photonic qubits. Each photonic qubit is of course formed from two states of polarization. We would
thus use two single photons propagating in opposite directions so as to keep the qubits
distinguishable. Note that single photons have been used extensively as qubits for quantum
computation based on linear optics \cite{fransonQC,dowling}. Further, the implementation of the
algorithm requires a number of Hadamard transformations of photons which can be done by using
quarter-wave and half-wave plates. We show in Fig. \ref{fig3} a possible experimental set-up for
the interaction of distinguishable photonic qubits. We use a degenerate optical parametric
amplifier to produce two photons simultaneously \cite{kwait95,franson,teich}. These photons
interact with an ensemble of atoms. The experimental scheme shows how to make the photons distinct.
The Deutsch-Jozsa algorithm requires that the polarization of two photons are measured
simultaneously. For this purpose, we propose use of single-photon detectors \cite{milburn}. The
photon ``1" (``2") can be detected in either of the detectors HD1 (HD2) and VD1 (VD2), which
detects it in either horizontal or vertical polarization. The polarization beam splitter before the
detectors preselects the polarization. Clearly, if the two detectors (HD1 or VD1 and HD2 or VD2)
click simultaneously, the polarization of photons can be measured and thus the experiment leads to
a conclusion for the algorithm.

\begin{figure}
\centerline{\scalebox{0.6}{\includegraphics{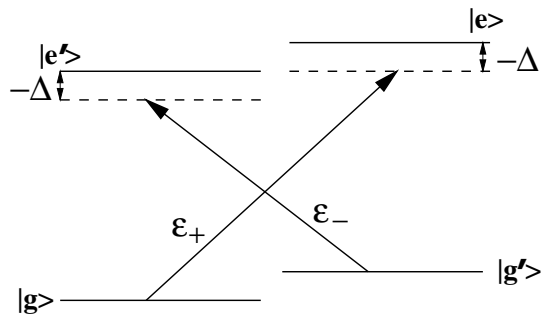}}} \caption{\label{fig2}Relevant level
configuration for implementing the Deutsch-Jozsa algorithm. The excited levels $|e\rangle$ and
$|e'\rangle$ are coupled to the ground levels $|g\rangle$ and $|g'\rangle$, respectively, by the
orthogonal polarization modes $\epsilon_+$ and $\epsilon_-$ of the photons, with equal detuning
$-\Delta$.}
\end{figure}

We consider the interaction of two distinct photons with an ensemble of $N$ atoms where only four
levels are relevant as shown in Fig. \ref{fig2}. Note that this kind of configuration can be found
in the clock transitions in $^{133}$Cs atoms and has been studied very extensively in
\cite{expts1,expts2}. Each photon has two circular polarization modes $\epsilon_\pm$. We designate
these modes as two orthogonal states $|1,0\rangle_k$ and $|0,1\rangle_k$, for the $k$th photon,
where $|1,0\rangle$ ($|0,1\rangle$) refers to a single photon in $\epsilon_+$ ($\epsilon_-$) mode
and no photon in $\epsilon_-$ ($\epsilon_+$) mode. The photons in the states $|1,0\rangle$ and
$|0,1\rangle$ interact respectively with the $|e\rangle\leftrightarrow |g\rangle$ and
$|e'\rangle\leftrightarrow |g'\rangle$ transitions of each atom. We assume that the common
frequency difference $\Delta$ between the photon polarization mode and atomic transition is much
larger than the atom-photon coupling constant $g$, which is the same for both the modes. In this
regime, the ground states $|g\rangle$ and $|g'\rangle$ of each atom get Stark-shifted. Using
second-order perturbation theory, the respective shifts can be calculated as $\hbar|g|^2n_+/\Delta$
and $\hbar |g|^2n_-/\Delta$, where $n_\pm$ are the total number of photons in $\epsilon_\pm$ modes.
As the excited states $|e\rangle$ and $|e'\rangle$ of the atoms remain unpopulated by interaction
with the photons in large detuning regime, the effective Hamiltonian can be written as
\begin{equation}
\label{effH1}H_{\mathrm{eff}}=\hbar\lambda\sum_{j=1}^N\sum_{k=1}^2\left[|g_j\rangle\langle
g_j||1,0\rangle_k\langle 1,0|+|g'_j\rangle\langle g'_j||0,1\rangle_k\langle
0,1|\right]\;,\;\lambda=|g|^2/\Delta\;.
\end{equation}
We rewrite the above Hamiltonian in terms of the horizontal ($|0\rangle_k$)and vertical
($|1\rangle_k$) polarization states of each $k$th photon as
\begin{eqnarray}
H_{\mathrm{eff}}&=&\frac{1}{2}\hbar\lambda\sum_{j=1}^N\sum_{k=1}^2\left[|g_j\rangle\langle
g_j|(|0\rangle_k\langle 0|-|1\rangle_k\langle 1|+i|0\rangle_k\langle 1|+i|1\rangle_k\langle
0|)\right.\nonumber\\ \label{effH}&+&\left.|g'_j\rangle\langle g'_j|(|0\rangle_k\langle
0|-|1\rangle_k\langle 1|-i|0\rangle_k\langle 1|-i|1\rangle_k\langle 0|)\right]\;,
\end{eqnarray}
where $|0\rangle_k=\frac{1}{\sqrt{2}}(|1,0\rangle_k+|0,1\rangle_k)$ and
$|1\rangle_k=\frac{1}{i\sqrt{2}}(|1,0\rangle_k-|0,1\rangle_k)$ are identified as two basis states
of the $k$th photonic qubit.  As we deal with only two states of each atom, the ensemble of $N$
atoms can have $2^N$ states. We use two of them as the states of the atomic qubit, namely
$\prod_{j=1}^N|g\rangle_j (\equiv |0\rangle_A) $ and $\prod_{j=1}^N|g'\rangle_j (\equiv
|1\rangle_A)$. Note that as the atomic qubit consists of only the ground states of the atoms,
decoherence due to spontaneous emission does not affect the process. Further, as the photons in
free space can have long decay time, they do not lead to any effective decoherence as well.

Under the action of the Hamiltonian (\ref{effH}), the following evolution occurs:
\begin{eqnarray}
|0\rangle_A|0,0\rangle_F &\longrightarrow&  -e^{-i\lambda
 Nt}|0\rangle_A(-i|0\rangle+|1\rangle)^{\otimes 2}\;,\nonumber\\
|0\rangle_A|1,1\rangle_F &\longrightarrow&e^{-i\lambda Nt}
|0\rangle_A(-i|0\rangle+|1\rangle)^{\otimes 2}\;,\nonumber\\
|0\rangle_A|0,1\rangle_F &\longrightarrow& -ie^{-i\lambda Nt}
|0\rangle_A(-i|0\rangle+|1\rangle)^{\otimes 2}\;,\nonumber\\
\label{phases}|0\rangle_A|1,0\rangle_F &\longrightarrow&-ie^{-i\lambda Nt}
|0\rangle_A(-i|0\rangle+|1\rangle)^{\otimes 2}\;,
\end{eqnarray}
where the state $|0,1\rangle_F$ represents the photon 1 in the horizontal polarization state
$|0\rangle_F$ and the photon 2 in the vertical polarization state $|1\rangle_F$ and so on. Similar
evolution occurs when the atomic qubit is initially in the state $|1\rangle_A$. Note that the
Hamiltonian (\ref{effH}) changes the polarization of each photon from linear to circular
polarization, i.e., the atomic ensemble acts as a polarizer for each photon. We next show how this
change in polarization helps us to implement the Deutsch-Jozsa algorithm for two qubits. Now
onwards, we choose the interaction time $T$ of the atomic ensemble with the photons such that
$\lambda NT=\pi/2$. As shown previously \cite{chandan}, this interaction time is achievable with
the available technology \cite{expts1,expts2}. For example, in an ensemble of $^{133}$Cs atoms in a
cell of dimensions 10$\times$10$\times$200 $\mu$m$^3$ at room temperature, the coupling constant
$g$ of each photon with the ensemble becomes $2.91\times 10^8$ s$^{-1}$ and time $T$ of interaction
of the photon with the ensemble becomes $6.67\times 10^{-13}$ s. Then for an atomic density of the
order of $5\times 10^{12}$ cm$^{-3}$ (i.e., for $N=10^5$), the required detuning of each photon can
be calculated from the above relation as $\Delta=3.59$ GHz $\approx 12.35g$. On the other hand, for
$D_1$ transitions (5$^2S_{1/2}\leftrightarrow 5^2P_{1/2}$, transition wavelength $\sim$ 795 nm) in
an ensemble of $^{87}$Rb atoms in a magneto-optical trap (with diameter $\sim$ 0.5 mm) \cite{lin},
we calculate $g\sim 3.53 \times 10^6$ s$^{-1}$ and $T\sim 1.67 \times 10^{-12}$ s. Thus for $N\sim
2.5\times 10^6$ in the trap, we find $\Delta\sim 9g$ Clearly, the condition $\Delta\gg g$ is well
satisfied in both cases.

We now introduce the following four inequivalent Hadamard operations for each photon in
$(|0\rangle_k, |1\rangle_k)$ basis and for each atom in the ensemble in ($|g\rangle_j,
|g'\rangle_j$) basis:
\begin{eqnarray}
h_1=\frac{1}{\sqrt{2}}\left(\begin{array}{cc}1&-1\\
1&1\end{array}\right)\;&;&
h_2=\frac{1}{\sqrt{2}}\left(\begin{array}{cc}1&i\\ i&1\end{array}\right)\;;\nonumber\\
\label{Hadamards}h_3=\frac{1}{\sqrt{2}}\left(\begin{array}{cc}1&1\\
-1&1\end{array}\right)\;&;&
h_4=\frac{1}{\sqrt{2}}\left(\begin{array}{cc}1&-i\\
-i&1\end{array}\right)\;.
\end{eqnarray}
These operations can be implemented by using a resonant microwave
field coupling the levels $|g\rangle_j$ and $|g'\rangle_j$ of the
$j$th atom. The corresponding interaction Hamiltonian is given by
\begin{equation}
\label{microH}
H^j_\mu=-\hbar\Omega\left[e^{i\phi}|g'\rangle_j\langle
g|+\textrm{H.c.}\right]\;,
\end{equation}
where $\Omega$ is the coupling constant of the microwave field with each atom in the ensemble and
$\phi$ is the phase of the field with respect to the matrix element between the levels
$|g\rangle_j$ and $|g'\rangle_j$. Under the action of this Hamiltonian, the rotations
(\ref{Hadamards}) for each atom can be obtained by choosing $\Omega t=\pi/4$ and $\phi=-\pi/2, 0,
\pi/2, \pi$, respectively. Note that we do not provide the above rotations for the collective
atomic states $|0\rangle_A$ and $|1\rangle_A$. Alternatively, one could use two classical fields in
Raman resonance to provide the equivalent rotation for each atom in the ensemble. For the photonic
qubits, it is easy to verify that the operations (\ref{Hadamards}) refer to rotation of linear
polarizations to linear or circular polarizations, which can be performed using different sequences
of quarter-wave and half-wave plates. Note that the matrix forms for transformations through
quarter-wave plates and half-wave plates are given by
\begin{eqnarray}
Q_\phi &=&
\frac{i}{\sqrt{2}}\left(\begin{array}{cc}\cos(2\phi)-i&\sin(2\phi)\\
\sin(2\phi)&-\cos(2\phi)-i\end{array}\right)\;;\nonumber\\
H_\phi &=& i\left(\begin{array}{cc}\cos(2\phi)&\sin(2\phi)\\
\sin(2\phi)&-\cos(2\phi)\end{array}\right)\;,
\end{eqnarray}
in $(|0\rangle_k, |1\rangle_k)$ basis, where $\phi$ is the angle of alignment of the wave-plate
with the axis perpendicular with its plane. The single-qubit operations (\ref{Hadamards}) for
photons can be obtained using a series of these wave-plates of different orientations [``the SU(2)
gadget"] as follows \cite{gsa_jmo,simon_mukunda}:
\begin{eqnarray}
h_1 \equiv Q_{\pi/4}Q_{\pi/4}H_{-3\pi/8}\;\;;\;\;h_2\equiv Q_{\pi/4} 
\nonumber\\
h_3\equiv Q_{\pi/4}Q_{\pi/4}H_{-\pi/8}\;\;;\;\;h_4\equiv Q_{-\pi/4}\;.
\end{eqnarray}
Note that the half-wave plate rotates the polarization state from $|0\rangle_k$ to $|1\rangle_k$
and vice versa for $\phi=\pi/4$.

Further, as required by the Deutsch-Jozsa algorithm, two photonic qubits should be initially in the
same polarization state $|0\rangle_k$. This can be done by using a half-wave plate [see Fig.
\ref{fig3}]. The degenerate parametric amplifier produces two photons in horizontal and vertical
polarization states. One of them (``2") is in vertical polarization $|1\rangle$ and thus is
reflected by a polarization beam splitter. This photon is then sent through a half-wave plate (HWP
in Fig.~\ref{fig3}) that rotates its polarization from $|1\rangle$ to $|0\rangle$, while the other
photon (``1"), which is already in horizontal polarization state $|0\rangle$, passes through the
polarization beam splitter. Thus both the photons are initialized to the state $|0\rangle_k$.

We now discuss how one can implement the Deutsch-Jozsa algorithm for two qubits using the evolution
(\ref{phases}) and (\ref{Hadamards}). We assume that two photons (the atomic ensemble) serve the
purpose of the system A (B) in Fig. \ref{fig1}. We start with the state
$|\psi_0\rangle_{F+A}=|0,0\rangle_F|1\rangle_A$ of the photons+ensemble system. Applying the
Hadamard rotation $h_1$ on each photon and the atomic qubit, we obtain
\begin{equation}
|\psi_1\rangle_{F+A}=\left[\frac{1}{\sqrt{2}}(|0\rangle+|1\rangle)\right]_F^{\otimes
2}\prod_{j=1}^N\frac{1}{\sqrt{2}}(|g_j\rangle-|g'_j\rangle)\;.
\end{equation}
Next the $U_{f_i}$ operations are applied. We show that in the
present case, $U_{f_i}=h_i^{\mathrm{eq}}U_{\mathrm{eff}}h_1^A$,
where $U_{\mathrm{eff}}=\exp[-iH_{\mathrm{eff}}t]$ is the unitary
operator corresponding to the Hamiltonian (\ref{effH}). Here the
operations $h_i^{\mathrm{eq}}$ correspond to different functions
$f_i(x)$.

In course of the operator sequence $U_{f_i}$, the operation
$h_1^A$ first prepares the atoms in $|0\rangle_A$ state:
\begin{equation}
|\psi'_1\rangle_{F+A}=\left[\frac{1}{\sqrt{2}}(|0\rangle+|1\rangle)\right]_F^{\otimes
2}|0\rangle_A\;.
\end{equation}
Then the operation $U_{\mathrm{eff}}$ for an interaction time $T$ defined by $\lambda NT=\pi/2$
yields the following [see Eqs. (\ref{phases})]:
\begin{equation}
\label{psi1DP}|\psi''_1\rangle_{F+A}\equiv\left[\frac{1}{\sqrt{2}}(-i|0\rangle+|1\rangle)\right]_F^{\otimes
2}|0\rangle_A\;.
\end{equation}
We next identify the operations $h_i^{\mathrm{eq}}$ for different
possible balanced functions $f_i(x)$ [see Table \ref{table1}] as
follows:
\begin{equation}
h_{3,4}^{\mathrm{eq}}\equiv
h''_{F_1}h'_{F_2}\;,~~h_{5,6}^{\mathrm{eq}}\equiv
h'_{F_1}h''_{F_2}\;,~~h_{7,8}^{\mathrm{eq}}\equiv
h''_{F_1}h''_{F_2}\;.
\end{equation}
Here $h'=h_1h_4h_3$ and $h''=h_1h_2h_3$ are composite sequences of
different Hadamard rotations defined in Eq. (\ref{Hadamards}) and
yield the following:
\begin{eqnarray}
&&h'|0\rangle=e^{i\pi/4}|0\rangle\;,\;h'|1\rangle=e^{-i\pi/4}|1\rangle\;;\nonumber\\
&&h''|0\rangle=e^{-i\pi/4}|0\rangle\;,\;h''|1\rangle=e^{i\pi/4}|1\rangle\;.
\end{eqnarray}
Note that $h'_{F_{i}}$ refers to $h'$ operations on $i$th photon.
The operators $h_i^{\mathrm{eq}}$ lead to the following results,
when applied on the state $|\psi''_1\rangle$ in Eq.
(\ref{psi1DP}):
\begin{eqnarray}
|\psi_2\rangle_{f_3,f_4}&=&h_{3,4}^{\mathrm{eq}}|\psi''_1\rangle\nonumber\\
&=&\pm\frac{1}{2}(|0,0\rangle+|0,1\rangle-|1,0\rangle-|1,1\rangle)_F|0\rangle_A\nonumber\\
&=&\pm\frac{1}{\sqrt{2}}(|0\rangle-|1\rangle)_{F_1}\frac{1}{\sqrt{2}}(|0\rangle+|1\rangle)_{F_2}|0\rangle_A\\
|\psi_2\rangle_{f_5,f_6}&=&h_{5,6}^{\mathrm{eq}}|\psi''_1\rangle\nonumber\\
&=&\pm\frac{1}{2}(|0,0\rangle-|0,1\rangle+|1,0\rangle-|1,1\rangle)_F|0\rangle_A\nonumber\\
&=&\pm\frac{1}{\sqrt{2}}(|0\rangle+|1\rangle)_{F_1}\frac{1}{\sqrt{2}}(|0\rangle-|1\rangle)_{F_2}|0\rangle_A\\
|\psi_2\rangle_{f_7,f_8}&=&h_{7,8}^{\mathrm{eq}}|\psi''_1\rangle\nonumber\\
&=&\pm\frac{1}{2}(|0,0\rangle-|0,1\rangle-|1,0\rangle+|1,1\rangle)_F|0\rangle_A\nonumber\\
&=&\pm\frac{1}{\sqrt{2}}(|0\rangle-|1\rangle)_{F_1}\frac{1}{\sqrt{2}}(|0\rangle-|1\rangle)_{F_2}|0\rangle_A\;.
\end{eqnarray}

The final step of the algorithm is to apply the Hadamard rotation
$h_1^F$ on each photon, which leads to the following outcomes:$
|\psi_3\rangle_{f_3,f_4}$$\equiv$ $|0,1\rangle_F|0\rangle_A$, $
|\psi_3\rangle_{f_5,f_6}$$\equiv$ $|1,0\rangle_F|0\rangle_A$, and
$|\psi_3\rangle_{f_7,f_8}$$\equiv$ $|0,0\rangle_F|0\rangle_A$.
Thus, after the final Hadamard rotations, either at least one of
the photons remains in $|0\rangle_F$ state, if the functions were
balanced.

On the other hand, in case of constant function $f_1$, the $U_{f_1}$ operation is equivalent to the
identity operation, while for the function $f_2$, the $U_{f_2}$ operation is equivalent to the NOT
operation on the atomic qubit. This can be implemented using microwave field for $\Omega t=\pi/2$
and $\phi=\pi/2$ [see the Hamiltonian (\ref{microH})]. Then, the output states
$|\psi_3\rangle_{f_1,f_2}$ of the photons+ensemble system become
$|1,1\rangle_F\prod_{j=1}^N\frac{1}{\sqrt{2}}(|g\rangle_j\pm |g'\rangle_j)$. This means that if the
functions were constant, both the photons can be detected in the state $|1\rangle_F$. Clearly, by
measuring the polarization states of two photons at the end of the algorithm, one can characterize
whether the functions were constant or balanced.

To this end, we discuss possibility of decoherence in the entire process. As the photons interact
with the ensemble dispersively, the excited states of the atoms get hardly populated. So the effect
of spontaneous emission during the process is minimal. However, there could be collisional
relaxation of the atomic ground states, which leads to loss of coherence in these states. But the
time-scales of both these relaxation processes are of the order of $10^{-6}$ s, whereas,
interaction time $T$ of the photon with the ensemble is of the order of $10^{-13}$ s. Clearly, this
is much less than the time at which the effects of different relaxation processes set in
significantly. Thus, our model with photons and atomic ensemble is virtually decoherence-free.

In conclusion, we demonstrated how the Deutsch-Jozsa algorithm can be implemented for two qubits
using an atomic ensemble. One requires two distinguishable photons for this. We show how this can
be achieved by using the output of a degenerate parametric amplifier. These photons dispersively
interact with an ensemble of $N$ atoms. The resulting Stark shifts of the atomic states lead to an
effective Hamiltonian, which along with several Hadamard transformations, are used to implement the
algorithm. We provide all the relevant operations to implement the algorithm and discussed the
expected outcomes.

One of us (G.S.A.) gratefully acknowledges the support from NSF grant No CCF 0524673.

\end{document}